\documentclass{jfm}
\usepackage{graphicx}
\usepackage{natbib}
\usepackage{color}
\usepackage{todonotes}
\usepackage{latexsym}
\usepackage{subfig}
\usepackage{natbib}
\usepackage{upgreek}
\usepackage{mathrsfs}
\usepackage{float}
\usepackage{caption}
\usepackage{soul}
\usepackage{textcomp}
\usepackage{stmaryrd}
\usepackage{amsmath}

\ifCUPmtlplainloaded \else
  \checkfont{eurm10}
  \iffontfound
    \IfFileExists{upmath.sty}
      {\typeout{^^JFound AMS Euler Roman fonts on the system,
                   using the 'upmath' package.^^J}%
       \usepackage{upmath}}
      {\typeout{^^JFound AMS Euler Roman fonts on the system, but you
                   dont seem to have the}%
       \typeout{'upmath' package installed. JFM.cls can take advantage
                 of these fonts,^^Jif you use 'upmath' package.^^J}%
      }
  \else
  \fi
\fi

\ifCUPmtlplainloaded \else
  \checkfont{msam10}
  \iffontfound
    \IfFileExists{amssymb.sty}
      {\typeout{^^JFound AMS Symbol fonts on the system, using the
                'amssymb' package.^^J}%
       \usepackage{amssymb}%
         
         \let\geq=\geqslant
      }{}
  \fi
\fi

\ifCUPmtlplainloaded \else
  \IfFileExists{amsbsy.sty}
    {\typeout{^^JFound the 'amsbsy' package on the system, using it.^^J}%
     \usepackage{amsbsy}}
    {\providecommand\boldsymbol[1]{\mbox{\boldmath $##1$}}}
\fi

\newsavebox{\astrutbox}
\sbox{\astrutbox}{\rule[-5pt]{0pt}{20pt}}

\def\ee{{\rm e}}
\def\ii{{\rm i}}

\def\der#1#2{{\partial #1\over \partial #2}}
\def\ddt#1{{D #1\over Dt}}
\def\ol#1{\overline{#1}}
\def\be{\begin{equation}}
\def\ee{\end{equation}}
\def\bea{\begin{eqnarray}}
\def\eea{\end{eqnarray}}
\def\bse{\begin{subequations}}
\def\ese{\end{subequations}}
\def\bsea{\begin{subeqnarray}}
\def\esea{\end{subeqnarray}}
\def\({\left (}
\def\){\right )}
\def\[{\left [}
\def\]{\right ]}
\def\<{\left <}
\def\>{\right >}

\newcommand{\qz}{\mbox{$\overline{q}_z$}}
\newcommand{\Uz}{\mbox{$\overline{U}_z$}}
\newcommand{\bz}{\mbox{$\overline{b}_z$}}

\affiliation{
$^1$ Department of Geosciences, Tel Aviv University,
Tel Aviv 69978, Israel.\\
[\affilskip]
$^2$ Mechanical Engineering Department, Indian Institute of Technology Kanpur, U.P. 208016, India.
}
\pubyear{2012}
\volume{xxx}
\pagerange{xxx--yyy}
\title{{A generalized action-angle representation of wave interaction in stratified shear flows}}
\author[E.~Heifetz and A.~Guha]
{
E\ls Y\ls A\ls L\ns H\ls E\ls I\ls F\ls E\ls T\ls Z$^{1}$% 
\and
A\ls N\ls I\ls R\ls B\ls A\ls N\ns G\ls U\ls H\ls A$^{2}$\footnote{Electronic mail for correspondence: anirbanguha.ubc@gmail.com }
}
\date{?? and in revised form ??}

\begin{document}

\maketitle

\begin{abstract}

In this paper we express the linearized dynamics of interacting interfacial waves in stratified shear flows in the  compact form of  action-angle Hamilton equations. The pseudo-energy serves as the Hamiltonian of the system, the action coordinates are the contribution of the interfacial waves to the wave-action, and the angles are their phases. The term ``generalized action-angle'' aims to emphasize that the action of each wave is generally time dependent and this allows instability. An attempt is made to relate this formalism to the action at a distance resonance instability mechanism between counter-propagating vorticity waves via the global conservations of pseudo-energy and pseudo-momentum. 

\end{abstract}

\section{Introduction}

Shear instability is a generic central phenomenon in fluid dynamics that has been extensively investigated since the end of the nineteenth century. Nevertheless, a simple intuitive understanding of the mechanism behind this instability is far from being straightforward. This stands in contrast, for instance, with thermal instability for which the basic understanding, that a heavy fluid above a lighter one tends to be unstable, agrees with our intuition and daily life experience. Furthermore, the essence of thermal instability can be understood in terms of the increasing offset of a parcel from its initial position, similar to a ball that is being pushed from a top of a hill and accelerates downward.
Shear flows do not provide such an immediate intuition; hence, whether a given shear flow setup has a tendency to become unstable cannot be concluded a-priori from physical arguments. In fact, there are setups which are apparently counter-intuitive, e.g. Taylor-Caulfield instability \cite[]{tayl1931,caul1995}. In some cases we may use mathematical constrains providing necessary conditions for instability, like the ones of Rayleigh, Fj\o rtoft and Richardson \cite[]{draz1982}. However these conditions do not provide a mechanistic understanding.

\begin{figure}
\centerline{\includegraphics[width=32pc]{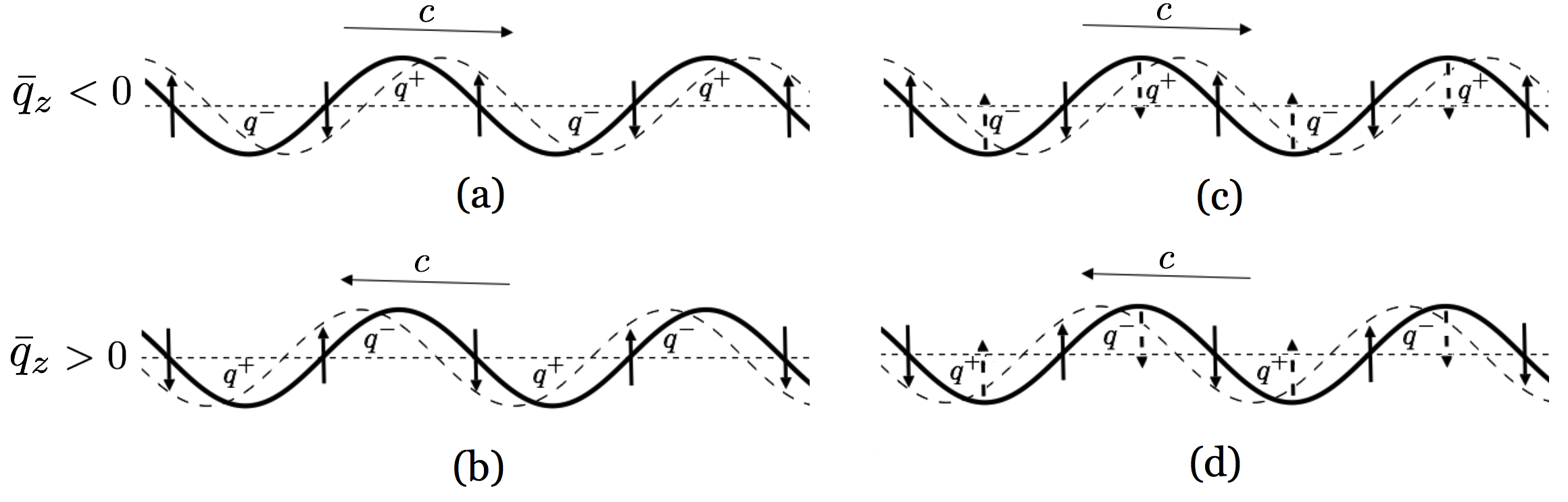}}
\caption{Two basic mechanisms of vorticity wave propagation. (a,b) Rossby waves: These occur in vorticity conserved flows satisfying $q = -\zeta \overline{q}_z$. Here $q$ gets generated from the background (potential) vorticity gradient $\overline{q}_z$. (c,d) Gravity waves: Here $q$ gets generated by the buoyancy restoring force, a quarter of wavelength phase-shifted to the right of the displacement field $\zeta$. In all sub-figures solid lines and arrows represent current snapshots of the waves whereas dashed lines and arrows represent the wave evolution from that current stage. Undulating lines represent the cross-stream displacement $\zeta$, vertical arrows represent the vertical velocity $w$ so that positive $\partial w/\partial x$ contributes to positive vorticity anomaly $q$. Horizontal lines represent the direction of wave propagation, relative to the local mean flow $\overline{U}$. }\label{fig:1abcd}
\end{figure}

\begin{figure}
\centerline{\includegraphics[width=32pc]{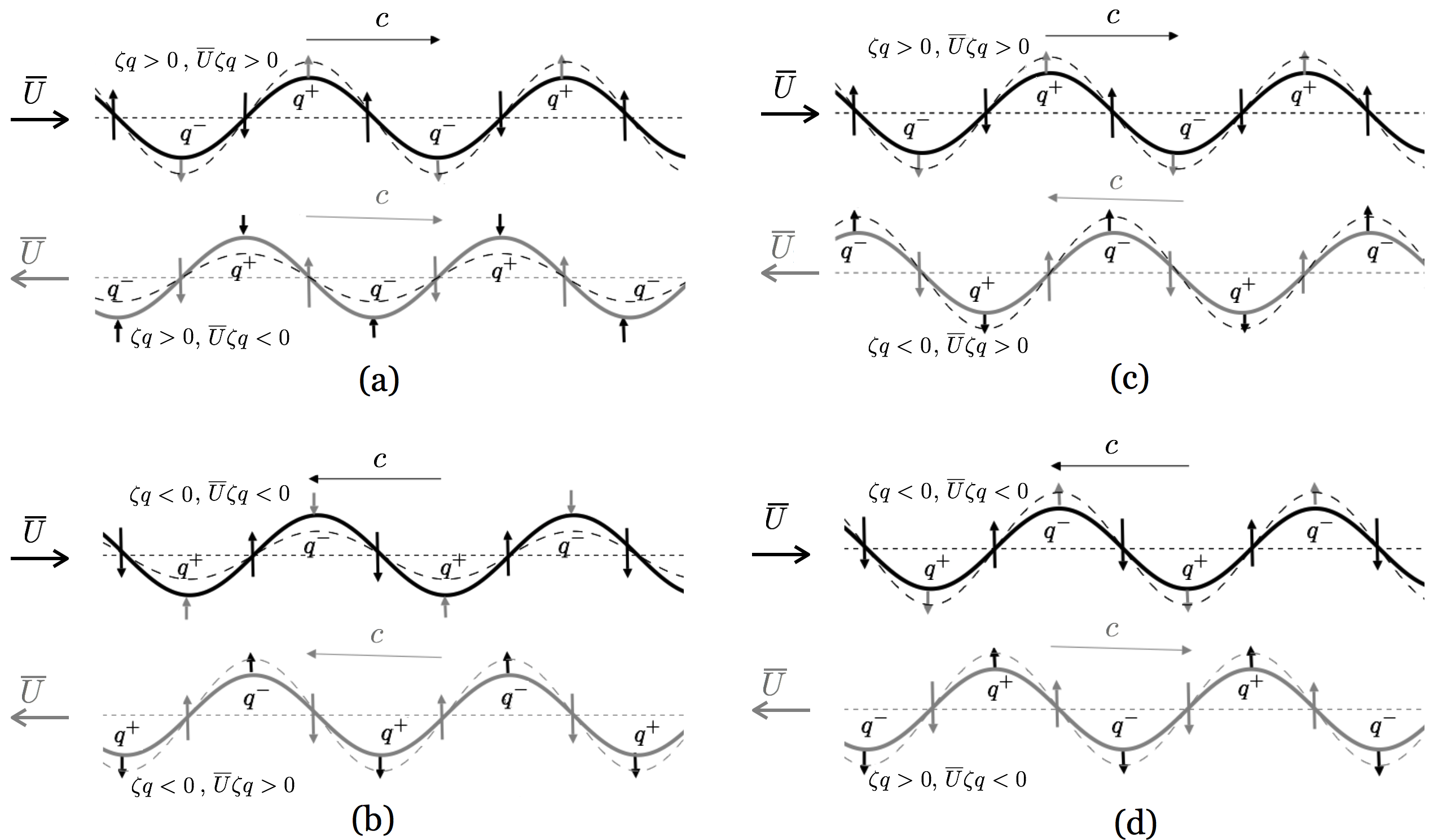}}
\caption{Two interfacial waves in presence of a background velocity shear; the latter is indicated by oppositely directed $\overline{U}$ at the two interfaces. Four cases are considered (a) pro-counter (leads to growth in one and decay in the other), (b) counter-pro (leads to decay in one and growth in the other), (c) pro-pro (leads to mutual instantaneous growth, which cannot be sustained), and (d) counter-counter (leads to sustained mutual growth and potential modal instability). }\label{fig:2abcd}
\end{figure}

In an attempt to develop a conceptual understanding of linear shear instability, a growing body of literature describes the instability in terms of resonant interaction at a distance between counter-propagating vorticity waves \cite[]{holm1962,bret1966,bain1994,caul1994,heif1999,heifetz2004counter,heif2005,carp2012,guha2014}. The core of the idea is illustrated in figures \ref{fig:1abcd}-\ref{fig:2abcd}. Let us consider for simplicity a 2D $(x-z)$ plane with a basic shear flow $\overline{U}(z)$ in the 
$x$ direction and define positive (negative) vorticity anomalies $q$, as resulting from counterclockwise (clockwise) anomaly circulations in this plane. 
From figure \ref{fig:1abcd} it is clear that a vorticity wave\footnote{We define any linear interfacial wave that propagates due to vorticity anomalies across the interface as a vorticity wave. Hence Rossby waves, gravity waves, capillary waves, Alfv\'en waves are all vorticity waves by our definition. } will be propagating to the right (left), relative to the local mean velocity $\overline{U}$, if its vorticity field is in phase (anti-phased) with its cross-stream displacement $\zeta$. In other words, $\zeta q>0$ implies a right moving wave, while $\zeta q<0$ implies a left moving wave. 

While the cross-stream velocity associated with the vorticity anomaly shifts the wave displacement, an additional mechanism is required to translate the vorticity anomalies in concert. For vorticity conserved flows, this could be the advection of the mean vorticity by the cross-stream velocity anomalies. This is the basic mechanism of Rossby  waves propagation, satisfying $q = -\zeta \overline{q}_z$, where $\overline{q}_z = -\overline{U}_{zz}$ is the basic state vorticity gradient (playing an equivalent role to the $\beta$ effect for planetary Rossby waves). 
Hence, the sign of $\overline{q}_z$ determines the direction of propagation: for negative (positive) values of $\overline{q}_z$ the waves propagate to the  right (left) with respect to the local mean velocity $\overline{U}$ (see figures \ref{fig:1abcd}(a,b)). For non-conserved vorticity flows a different basic mechanism to propagate the vorticity anomaly may result from the restoring force acting on the wave displacement. In this paper we will consider only a stably stratified configuration in which buoyancy acts to restore fluid parcels back to their initial positions. As illustrated in figures \ref{fig:1abcd}(c,d), the vertical motions
associated with this restoring force generate horizontal shear anomalies ($\partial w/\partial x$) and thus a vorticity field $q$. This baroclinic vorticity generation is phase shifted by a quarter of wavelength to the right of the displacement field $\zeta$. Therefore, in both cases of propagation, whether to the right or to the left, the translation of $q$ is in concert with $\zeta$.

While each vorticity wave in isolation is neutral, instability may result from interaction between the waves. The interaction is mediated by the far field velocity that each wave induces on the other. An instantaneous mutual amplification may be obtained when the induced cross-stream velocity by each wave is in phase with the cross-stream displacement of the other one. In such a configuration, fluid parcels of the two waves are pushed further away from their initial rest positions, signifying instability. In figure \ref{fig:2abcd} we sketch four possible characteristic snapshots of interactions. We can see that mutual amplification is possible only between pairs of waves with opposite $(\zeta,q)$ sign relations. Therefore, based only on this inspection, we may expect the possibility of instability when the domain integration of the correlation between $\zeta$ and $q$ fields, i.e.\ $\<\zeta q\>$ (angle brackets denote domain integration) is small, or even zero due to symmetry between mutually amplifying pairs of waves.  

Furthermore, in order to sustain such mutual amplification, the waves should be in a phase-locked configuration (then phase-locking and mutual growth may lead to normal mode instability). However, as discussed above, the $(\zeta,q)$ sign relation determines the direction of propagation. Thus, two waves with opposite sign relations will propagate in opposite directions, and will therefore fail to lock in phase. Nevertheless, the mutual growth configuration can be maintained in the presence of a mean shear, provided each wave propagates counter to its local mean flow (such waves are referred to as ``counter-propagating vorticity waves''). The different configurations for which mutual growth may or may not sustain are illustrated in figures \ref{fig:2abcd}(a-d). On inspecting these figures we may expect that the spatial correlation $\<\overline{U}\zeta q\>$ to be negative for sustained mutual growth. The reason can be explained as follows: $\zeta q >0$ implies a wave whose intrinsic propagation is rightward, and its propagation can only be hindered if $\overline{U}<0$. The opposite is true for the leftward propagating wave. Hence for counter-propagation, $\overline{U}\zeta q$ should be negative for both waves.

In this paper we intend to show that this conceptual understanding is imprinted in the conservation laws of pseudo-momentum (PM) (or wave-action (WA) for a given zonal wavenumber) and pseudo-energy (PE) (thorough derivations of PM and PE can be found in \cite{Buhler2009}), which are the two constants of motion for linearized stratified shear flows. Those resulted, respectively from the zonal symmetry and the time independence of the mean flow\footnote{The symmetry of the mean flow in the streamwise direction, as well as the steadiness of the mean flow in the linearized dynamics overcome the general intrinsic difficulty of particle relabeling symmetry that generally prevents canonical Hamiltonian formulation of fluid flows \cite[]{salmon1988,shepherd1990}.}. In fact, the condition for mutual wave amplification is derived from the vanishing of PM (or WA) for normal mode instability. Likewise, the condition for counter-propagation and hence phase-locking is derived from the vanishing of PE. Furthermore, we generalize the results obtained for vorticity conserved shear flows \cite[]{heifetz2009canonical} in order to accommodate the effects of density stratification, and show that the vorticity wave interaction equations translate to the {\emph{generalized}} action-angle (A-A) Hamilton's equations  (this generalization is discussed in detail in \S \ref{sec:gen_action_angle}). In these equations PE is the Hamiltonian, WA is the action, and the waves' phases serve as the angle coordinates\footnote{In this paper the formulation will be derived directly from the properties of the linearized wave dynamics. In standard classical mechanics, A-A is obtained from the generalized momenta and coordinates  $({q}_i, {p}_i)$, where  $i$ denotes a component of the action $J$ such that $J_i = \oint p_i dq_i$. In the context of linearized dynamics $i$ represents the zonal component of the circulation integral on constant density surfaces. However such derivation is somewhat out of the focus of this paper and therefore will not be presented here.}.

The paper is organized as follows. In \S \ref{sec:2}, we introduce the linearized stratified shear flow dynamics in a vertical slice model and derive the two constants of motion, PM and PE, in terms of $\zeta$ and $q$. Then, we discuss how these conservation laws agree with the paradigm of wave interaction. In \S \ref{sec:3}, we derive the WA and relate it to PM and PE and obtain the A-A formulation. First we recover the known relations for plane-waves in constant stratification and zero shear and next for interfacial waves in general shear and stratification. { In \S \ref{sec:4} we  explicitly show that the complex wave interactions for two interfaces can be compactly expressed as a generalized A-A formulation}, and finally discuss our results in \S \ref{sec:5}.

\section{Pseudo-momentum and pseudo-energy in stratified shear flows}
\label{sec:2}
\subsection{Linearized dynamics formulation}

%Hereafter we follow the notations of \cite{harnik2008} and denote that paper by BVW08.

We consider a Boussinesq, 2-D flow slice model in the zonal (stream-wise) vertical (cross-stream) plane ($x-z$), with a zonally uniform basic
state (denoted by over-bars) which varies with height and is in hydrostatic balance. The momentum and continuity equations, linearized around
this base state are:

\be
\ddt u=-w\Uz-\frac{1}{\rho_0}\der{p}{x},
\label{equmom}
\ee
%\addtocounter{equation}{-1}
%%\renewcommand{\theequation}{\arabic{equation}b}
\be
\ddt w=b-\frac{1}{\rho_0}\der{p}{z},
\label{eqwmom}
\ee
%\addtocounter{equation}{-1}
%%\renewcommand{\theequation}{\arabic{equation}c}
\be
\ddt b=-wN^2,
\label{qbouy}
\ee
%\addtocounter{equation}{-1}
%%\renewcommand{\theequation}{\arabic{equation}d}
\be
\der{u}{x}+\der{w}{z}=0
\label{eqnondiv}.
\ee
\noindent Here $D/Dt \equiv \partial/\partial t + \ol{U}\partial/\partial x$ is the linearized material derivative; %$\ddt{}\equiv\der{}{t}+\ol{U}\der{}{x}$
 ${\bf u}=(u,w)$ is the perturbation velocity vector; ${\overline U}(z)$ is the zonal mean flow; $p$ is the perturbation pressure; $\rho$ is the perturbation density; 
$\rho_0$ is a constant reference density and $b = -\rho g/\rho_0$ is the perturbation buoyancy. The squared buoyancy (or Brunt--V\"ais\"al\"a) frequency is defined as  $N^2 \equiv -(g/\rho_0)d\overline\rho/d z=\overline{b}_z$, where $\overline{\rho}$ is the mean density, $g$ denotes gravity, $\overline{b} \equiv -\overline\rho g/\rho_0$ is the mean buoyancy, and the subscript $z$ denotes the vertical derivative. 

% is the 
% Brunt--V\"ais\"al\"a frequency where $\rho$ and $\overline{\rho}$ are the perturbation and mean flow density, respectively, and $g$ denotes gravity. We note that $N^2=\overline{b}_z$, and use this in further notation.

Equation set (\ref{equmom})-(\ref{eqnondiv}) can be transformed straightforwardly into a single equation in terms of the perturbation vorticity %$q=\der{w}{x}-\der{u}{z}$
$q \equiv \partial w/\partial x - \partial u/ \partial z $ and vertical displacement $\zeta$ fields:
% %\renewcommand{\theequation}{\arabic{equation}} 
\be
%\ddt q=-w\qz+\der{b}{x}
{D \over Dt} (q+ \qz\zeta) = -\bz\der{\zeta}{x},
\label{eqq}
\ee
where $D\zeta/Dt =w$ (from the kinematic condition) and $\qz=-\overline{U}_{zz}$ is the mean vorticity gradient.
For homogeneous  fluids ($\bz = 0$) the LHS indicates that vorticity perturbation is generated by vertical advection of the mean vorticity (the Rossby mechanism, sketched in figures \ref{fig:1abcd}(a,b)), whereas in density stratified plane Couette flows (flows with constant shear, i.e.\ $\qz = 0$), the RHS indicates that vorticity is generated due to the buoyancy restoring mechanism illustrated in figures \ref{fig:1abcd}(c,d).   

\subsection{Pseudo-momentum and Pseudo-energy conservations}

We assume zonal boundary conditions to be periodic and the vertical velocity to vanish at the {upper and lower} horizontal boundaries. Hence the domain integration  of the cross-stream vorticity flux vanishes, i.e., $\< w q \> = 0$. Hence,
multiplying  (\ref{eqq}) by $\zeta$ and integrating by parts yields the conservation of pseudo-momentum ${\cal P}$: 
\be
\der{}{t}{\cal P} = 0,
\ee
where 
%\addtocounter{equation}{-1}
%\renewcommand{\theequation}{\arabic{equation}b} 
\be
{\cal P} \equiv \<\zeta\(q+{\qz\over 2} \zeta\)\>.
\label{eq:PM}
\ee 
For the homogeneous case of zero stratification $q = -\qz\zeta$, the familiar expression 
\be
{\cal P}_h = {1\over 2}\<\zeta q\> = -\<{q^2 \over 2 \qz}\> = -\<{\qz \over 2 }\zeta^2\>,  
\ee
is recovered and leads to the Rayleigh inflection point condition. For the stratified Couette like flow of constant shear we simply obtain
%\addtocounter{equation}{-1}
%\renewcommand{\theequation}{\arabic{equation}b} 
\be
{\cal P}_c = \<\zeta q\>.  
\ee
Thus, modal instability, for which ${\cal P}=0$, can be satisfied by pairs of waves with opposite $(\zeta, q)$ sign relations that mutually amplify each other in accordance with figures \ref{fig:2abcd}(c,d).  

Defining ${\it P} \equiv \zeta\(q+{\qz} \zeta/2\)$ as the integrand of ${\cal P}$, the pseudo-energy conservation is obtained from (\ref{equmom})-(\ref{eqnondiv}) when noting that
\be
\der{}{t}{\cal E} = -\<\ol{U} wq\> = -\der{}{t}\<\ol{U}{\it P}\>,
\ee
where ${\cal E} \equiv \<E\>$ is the sum of the eddy kinetic and available potential energies:
\be
\<E\> \equiv \<EKE\> + \<APE\>= \<{1\over 2}\(u^2+w^2\)\> +\<{\bz\over 2}\zeta^2\>.
\label{eq:En}
\ee
The pseudo-energy conservation then reads
\be
\der{}{t}{\cal H} = 0,
\ee
where 
%\addtocounter{equation}{-1}
%\renewcommand{\theequation}{\arabic{equation}b} 
\be
{\cal H} \equiv \<H\>; \hspace{0.5cm} H=E + \ol{U}{\it P}.
%{\cal H} \equiv \<E + \ol{U}{\it P}\>.
\label{eq:PE}
\ee 
For modal instability the pseudo-energy integral vanishes, therefore $\<\ol{U}{\it P}\> = -\<E\> < 0$. Hence, in both the homogeneous and the Couette like cases this implies that $\<\ol{U}\zeta q\> < 0$, which is the condition for counter-propagation (figure 2(c,d)). For the homogeneous case we obtain the familiar Fj\o rtoft condition $\< \ol{U} q^2  / (2\qz )\> > 0$ indicating positive correlation between the mean flow  $\ol{U}$ and the mean vorticity gradient $\qz$. For the more general setup of stratified shear flow we show in Appendix \ref{App:A} that the Howard-Miles criterion \citep{draz1982} for modal instability is related to the vanishing of PE. The simultaneous satisfaction of the two conditions of ${\cal P} = 0$ and ${\cal H} = 0$ are therefore in agreement with the conditions of mutual amplification and phase locking between pairs of vorticity waves (figure 2(d)).

\section{Wave-action, pseudo-momentum, pseudo-energy and the action-angle formalism}
\label{sec:3}

\subsection{Plane-waves in constant stratification and zero shear}

For the sake of simplicity let us first consider  the case of constant stratification 
and mean flow ($N^2 =\overline{b}_z = \mathrm{const}_1$, $\ol{U} = \mathrm{const}_2$). Equation set 
(\ref{equmom})-(\ref{eqnondiv}) admits the familiar plane wave solution of the form of $e^{-i\theta}e^{i(kx+mz)}$, where the phase: $\theta = \omega t + \theta_0$ ($k$ and $m$ are the zonal and vertical wave number components and $k$ is assumed positive, $\omega$ is the wave frequency and $\theta_0$ is the initial phase) with the dispersion relation 
\be
c= {\omega\over k} = \ol{U} + \hat{c}; \hspace{0.5cm} 
\hat{c} = {\hat{\omega}\over k} = \pm {N \over \sqrt{k^2 +m^2}},
\label{eq:pwdr}
\ee
where $c$ is the phase speed in the zonal direction and $\hat{\omega}$ and $\hat{c}$ denote the intrinsic frequency and phase speed in the reference frame of the mean flow. It is straightforward to show that  the zonal-averaged wave energy is equi-partitioned between its kinetic and potential counterparts  so that $\ol{E} = \overline{{b}_z \zeta^2}$ \citep{Buhler2009}. Defining the zonal-averaged wave action (here after WA) as 
$\ol{A} \equiv \ol{E}/\hat{\omega}$ {(note that action in classical mechanics is traditionally referred by the symbols $J$ or $I$, however in fluid mechanics, wave action is usually referred by the symbol $A$, see \cite{Buhler2009})} we obtain its relations with the zonal-averaged PM and PE:
\be 
\ol{P} = k\ol{A} = {\ol{E}\over\hat{c}}; \hspace{0.5cm} 
\ol{H} = \ol{E +UP} = \omega \ol{A} = {\omega \over \hat{\omega}}\ol{E}.
\label{eq:pwwa}
\ee
We note then that the sign of the contribution of the wave to PM is equal to the sign of the intrinsic frequency, whereas the contribution to PE is positive unless $\hat{c}$ has the opposite sign of $\ol{U}$ and $|\hat{c}| < |\ol{U}|$, that is when the wave propagates counter the mean flow, however with an intrinsic phase speed that is not large enough to  overcome the mean flow advection. 

From the second equation in (\ref{eq:pwwa}) we can deduce that
\be 
\ol{H} = \dot{\theta}\ol{A} \hspace{0.25cm} \Rightarrow \hspace{0.25cm}
\der{\ol{H}}{\ol{A}} = \dot{\theta},
\label{eq:pwaa1}
\ee
and since $\ol{H}$ is time independent 
\be 
\dot{\ol{H}} = \der{\ol{H}}{\ol{A}}\dot{\ol{A}} + 
\der{\ol{H}}{{\theta}}\dot{{\theta}}=0 \hspace{0.25cm} \Rightarrow \hspace{0.25cm}
\der{\ol{H}}{\theta} = -\dot{\ol{A}}.
\label{eq:pwaa2}
\ee
Equations (\ref{eq:pwaa1}) and (\ref{eq:pwaa2}) provide the canonical action ($\ol{A}$) -- angle ($\theta$) (hereafter A-A) representation of the plane wave linearized dynamics\footnote{As mentioned previously, in classical mechanics the action is obtained from the circulation integral 
$J = \oint {\bf u}d{\bf x}$. For stratified shear flow the circulation (of the mean flow plus perturbation) should be evaluated on an undulating plane of constant density where the baroclinic torque is zero and  circulation is conserved. Under the linearized wave dynamics \cite{Buhler2009} showed that the Eulerian representation of the zonal component of $J$ has a constant contribution from the initial steady mean flow and a contribution that is proportional to the negative of PM. The latter is the order $O(\epsilon^2)$ mean flow response. Hence $J$ is proportional to the negative of the wave action, so in principle we should define the wave action as $-\ol{A}$ and the angle as $-\theta$. However, since in the context of fluid dynamics wave action is defined as $\ol{A}$, we follow this convention in the text.}. Obviously in this simple case $\ol{H}$ is not a function of the wave phase (angle) and indeed $\ol{A}$ is time independent. This is the standard A-A formulation {in which action is a  constant of motion}. This simple example has been brought in order to pave the road for the \emph{generalized} A-A description of the interaction between interfacial vorticity waves in stratified shear flows, where the total action is conserved but the action of each wave is generally time dependent. 

\subsection{Single interfacial vorticity wave}
\label{subsec:3.2}
The presence of shear distorts the structure of plane waves. Thus, let us now consider interfacial waves that are resilient to shear and preserve their untilted structures.  
For a stratified shear flow, where both the mean vorticity and the stratification are discontinuous at some level $z=z_0$:
\be
\qz=\Delta\overline{q}_0\delta(z-z_0)\, ,
\hspace{0.5cm}
\bz=\Delta\overline{b}_0\delta(z-z_0),
\label{eq:SIBS}
\ee
where $\Delta\overline{q}_0 \equiv\overline{q}(z>z_0)-\overline{q}(z< z_0)$, and $\Delta\overline{b}_0 \equiv\overline{b}(z>z_0)-\overline{b}(z< z_0)$,
 (\ref{eqq}) then implies that the vorticity perturbation should  as well have the form of a $\delta$-function at $z=z_0$. Thus, for a normal mode solution we may write
\be
q= {\tilde q}_0e^{\ii k(x-c t)}\delta(z-z_0).
\label{eq:vorint}
\ee
We can find the wave dispersion relation and structure on combining (\ref{eqq}) with the kinematic condition ${D\zeta/ Dt}=w$ at $z=z_0$ and express $w$ in terms of $q$ via the Green's function formulation (e.g. \cite{harnik2008}): 
% The material derivative of (\ref{eqq}) yields: 
% %\renewcommand{\theequation}{\arabic{equation}}
% \be
% {D^2 q\over Dt^2 } = -\( \qz {Dw\over Dt} + \bz\der{w}{x}\),
% \label{eq:master}
% \ee
% which is traditionally converted into the Taylor-Goldstein equation by looking at a normal mode solution of the form of $e^{\ii k(x-ct)}$ and expressing both $w$ and $q$ in terms of the  streamfunction. Alternatively, we follow \cite{harnik2008} and consider a Fourier component solution of the form of $e^{\ii kx}$ to express $w$ in terms of $q$ via a Green's function formulation:
\be
w(z)=\int_{z'}q(z')G(z',z,k)dz',
\label{eqwqG}
\ee
where $-k^2G+\partial^2 G/\partial z^2=\ii k\delta(z-z')$. The Green's function  $G(z',z,k)$ depends on the boundary conditions and on the zonal wavenumber $k$. For open flows, which we assume here for simplicity (Different Green's functions for different boundary conditions can be found in \cite{heif2005})
\be 
G(z',z,k)=-\frac{\ii}{2}e^{-k|z-z'|},
\label{eqG}
\ee
 so that $w(z=z_0) = -\ii {\tilde{q}\over 2} e^{\ii k(x-c t)}$. The  dispersion relation obtained satisfies
\be
c^{\pm}= \ol{U}_0 + \hat{c}^{\pm}; \hspace{0.5cm} 
\hat{c}^{\pm} = -{\Delta\overline{q}_0\over 4k}\pm\sqrt{\({\Delta\overline{q}_0\over 4k}\)^2+{\Delta\overline{b}_0\over 2k}\;}.
\label{eq:SIDR}
\ee 
Thus for stable stratification ($\Delta\overline{b}_0 > 0$), $\hat{c}^+$ is always positive and $\hat{c}^-$ is always negative. The associated terms with $\Delta\overline{q}_0$ reflect the Rossby wave propagation mechanism (figure \ref{fig:1abcd}(a)-(b)) and the term with $\Delta\overline{b}_0$ results from the buoyancy restoring force (figure \ref{fig:1abcd}(c)-(d)). When $\Delta\overline{q}_0 = 0$ we recover the familiar deep water internal wave dispersion relation, whereas when $\Delta\overline{b}_0 = 0$ we recover the interfacial Rossby wave together with a degenerated solution of zero vorticity perturbation. The asymmetry between $\hat{c}^+$ and $\hat{c}^-$ results from the Rossby wave mechanism propagating the wave to the left of the mean vorticity gradient, whereas the buoyancy restoring force is even for both rightward and leftward propagation. The  two interfacial waves at the interface satisfy
\be
q= \[{\tilde q}_0^+e^{-\ii kc^+t} + {\tilde q}_0^-e^{-\ii kc^-t}\]\delta(z-z_0)e^{\ii kx}\, ,
\hspace{0.25cm}
\zeta(z=z_0) = \[\zeta_0^+e^{-\ii kc^+t} + \zeta_0^-e^{-\ii kc^-t}\]e^{\ii kx},
\label{eq:eigen_m}
\ee
where
\be
\tilde{q}_0^{\pm}=2k \hat{c}^\pm\zeta_0^\pm.
\label{eq:qzeta}
\ee
These are in agreement with figure \ref{fig:1abcd}, indicating that the wave whose vorticity and displacement structures are in (anti) phase propagates to the (left) right with respect to the local mean flow. 

Despite the presence of a mean shear (generally $\ol{q}= -\ol{U}_z(z) \neq 0$) these interfacial waves preserve their untilted structure throughout the domain.
Defining the perturbation streamfunction $\psi$, so that $u=-\psi_z$; $w=\psi_x$; $q=\nabla^2\psi$, and recall that $\<EKE\> = -\<\psi q\>/2$, the domain integrated wave energy for each wave can be evaluated solely from the interface using (\ref{eq:SIBS}), (\ref{eq:vorint}), (\ref{eqG}) and (\ref{eq:qzeta}) :
\be
{\cal E}^{\pm} = \[\<EKE\> + \<APE\>\]^{\pm}= 
 \[-\<{\psi q\over 2}\> +\<{\bz\over 2}\zeta^2\>\]^{\pm} = 
{k\over 2}\[\(\hat{c}^{\pm}\)^{2} + {\Delta \ol{b}_0 \over 2k}\]
|{\zeta_0}^{\pm}|^2,
\label{eq:sienergy}
\ee
where the shear prevents the equi-partition between the kinetic and the potential energies of the waves. Substitution of (\ref{eq:qzeta}) in (\ref{eq:sienergy}) indicates that the two interfacial waves are orthogonal to each other under the energy norm, i.e., 
$\<E\> = \<E\>^{+} +\<E\>^{-}$. This is due to the fact that the integrated energy results solely from the waves' signatures at the interface. Generally it is PE (which is the sum of the wave energy and the second order mean flow response) rather than the the wave energy itself, that is conserved under linearized dynamics. The contribution of each wave to PM is obtained by substituting the displacement and vorticity of the waves at the interface (i.e.\ $\tilde{q}_0^{\pm}$ and $\zeta_0^{\pm}$, and using (\ref{eq:qzeta})) in (\ref{eq:PM}): 
\be 
{\cal P}^{\pm} = \pm {1\over 2}|{\zeta_0}^{\pm}|^2\sqrt{\({\Delta\overline{q}_0\over 2}\)^2+2k\Delta\overline{b}_0\;}\, ; \hspace{0.25cm}  {\cal P} = {\cal P}^{+} +{\cal P}^{-}.
\label{eq:sinpe}
\ee
Thus, the contribution of the rightward (leftward) propagating interfacial wave to PE is positive (negative). Furthermore, as expected, neutral normal modes with different phase speeds are orthogonal with respect to a norm that is a conserved quantity \citep{held1985pseudomomentum}. 
Defining the domain integrated WA as 
${\cal A} \equiv {\cal E}/\hat{\omega}$, it is straightforward to verify for the interfacial waves that
\be 
{\cal A}^{\pm} = {{\cal P}^{\pm} \over k}\, ; \hspace{0.25cm}  
{\cal A} = {\cal A}^{+} +{\cal A}^{-}.
\label{eq:sinwa}
\ee
Hence the interfacial waves are orthogonal as well with respect to WA. To complete the analogy with plane-waves in absence of shear, we note as well that
\be 
{\cal H} = {\cal H}^+ + {\cal H}^- = \({\omega}{\cal A}\)^+ +  \({\omega}{\cal A}\)^- \, . 
\label{eq:sinpe}
\ee
Defining the phase angle for the interfacial waves $\theta^{\pm}$ to satisfy 
$\dot{\theta}^{\pm} =  {\omega}^{\pm}$, we obtain the canonical A-A formulation
\be 
{\cal H} = \(\dot{\theta}{\cal A}\)^+ + \(\dot{\theta}{\cal A}\)^- , 
\label{eq:sinaa1}
\ee
where
\be 
\der{\cal H}{{\cal A}^{\pm}} = \dot{\theta}^{\pm}\, ; \hspace{0.25cm}
\der{\cal H}{{\theta}^{\pm}} = -\dot{\cal A}^{\pm} =0\, .
\label{eq:sinaa2}
\ee

\subsection{Multiple interfaces}
\label{sec:gen_action_angle}

For the interfacial wave dynamics we discretize the continuous mean flow into piecewise linear profiles of vorticity and density (buoyancy): 
\be
\qz=\sum_{n=1}^{N} \Delta\overline{q}_n\delta(z-z_n)\, ,
\hspace{0.5cm}
\bz=\sum_{n=1}^{N} \Delta\overline{b}_n\delta(z-z_n),
\label{eqqzbzjumps}
\ee
where $\Delta\overline{q}_n=\overline{q}(z_{n+1}>z>z_n)-\overline{q}(z_n > z > z_{n-1})$, and $\Delta\overline{b}_n=\overline{b}(z_{n+1}>z>z_n)-\overline{b}(z_n > z > z_{n-1})$. 
This formulation may include interfaces with only density jumps, only vorticity jumps, or both. Thus, it may be applied to different basic setups such as Rayleigh, Holmboe and Taylor-Caulfield profiles.
In the next section we show that the linearized interfacial wave dynamics can be presented in the compact A-A form
\be 
{\cal H} = \sum_{n=1}^N\[\(\dot{\theta}{\cal A}\)^+ + \(\dot{\theta}{\cal A}\)^-\]_n \, ; \hspace{0.25cm}
{\cal A} = \sum_{n=1}^N\[{\cal A}^+ + {\cal A}^-\]_n ; \hspace{0.25cm}
\der{\cal{H}}{{\cal A}^{\pm}}_n = \dot{\theta}_n^{\pm}\, ; \hspace{0.25cm}
\der{\cal H}{{\theta}^{\pm}}_n = -\dot{\cal A}_n^{\pm}\, ,
\label{eq:multAA}
\ee
%\todo{(3.19) is canonical Hamiltonian form, not action angle form}
where ${\cal H}$ and ${\cal A} = k{\cal P}$ are the two constant of motion of the linearized monochromatic wave interaction dynamics. As opposed to (\ref{eq:sinaa2}) and to textbook examples in classical mechanics, the WA components, ${\cal A}_n^{\pm}$, are generally non-zero due to interaction at a distance between remote interfacial waves. {While it may be argued that the last two equations in (\ref{eq:multAA})  straight-forwardly indicates canonical Hamilton equations, we have however chosen to refer (\ref{eq:multAA})  as ``generalized A-A equations''. This is due to the additional constraint in our system - the sum of all the individual wave actions, $\cal A$ is constant.} 
%This allows instability and therefore we denote  (\ref{eq:multAA}) as ``generalized A-A dynamics''. 
 For modal instability of the form of  $e^{\ii[kx-(\omega_r + \ii\omega_{\ii}) t)]}$, 
${\cal H} = \omega_r {\cal A} =0$, however
%\todo{I am not sure whether reviewers will accept this as  action angle because by definition $\cal H=\cal H(\cal A)$ for $\theta$ to be an ``angle/cyclic'' coordinate. Furthermore, the Arnold-Liouville theory of integrability will also won't apply. }
\be 
\der{\cal H}{{\cal A}^{\pm}}_n = \dot{\theta}_n^{\pm} = \omega_r\, ; \hspace{0.25cm}
-{1 \over {\cal A}_n^{\pm}}\der{{\cal H}}{{\theta}^{\pm}_n} = 
\({\dot{\cal A}_n^{\pm}\over {\cal A}_n^{\pm}}\) = 2\omega_{\ii}\, .
\label{eq:multAANM}
\ee

\section{Explicit derivation of A-A for two interfaces wave interaction}
\label{sec:4}

In order to appreciate the compactness of (\ref{eq:multAA}), here we explicitly derive   the wave interaction equations for two interfaces. The generalization for multiple interfaces follows naturally.

\subsection{PM, Energy, WA, and PE for two interfaces}

Consider now two interfaces located at $z_1$ and $z_2 = (z_1 + \Delta z)$, so that $N=2$ in (\ref{eqqzbzjumps}). Let us decompose the displacement and the vorticity perturbations at those interfaces into the interfacial waves discussed in  \S \ref{subsec:3.2}:
\be
\nonumber
q_{1,2}= \[{\tilde q}^+(t)+{\tilde q}^-(t) \]_{1,2}\delta(z-z_{1,2})e^{\ii kx} = 
\[Q^+ (t) e^{-\ii \theta^+ (t)} + Q^- (t) e^{-\ii \theta^- (t)} \]_{1,2}\delta(z-z_{1,2})e^{\ii kx}\, ,
\ee
\be
\zeta(z=z_{1,2}) = \[{\tilde \zeta}^+(t)+{\tilde \zeta}^-(t) \]_{1,2}e^{\ii kx} = 
\[Z^+ (t) e^{-\ii \theta^+ (t)} + Z^- (t) e^{-\ii \theta^- (t)}\]_{1,2}e^{\ii kx},
\label{eq:eigen_m2int}
\ee
where 
\be
{\tilde q}^{\pm}_{1,2}= \[\(2k\hat{c}^\pm\) {\tilde \zeta}^\pm\]_{1,2},
\label{eq:qzeta2int}
\ee
and
\be
\hat{c}^{\pm}_{1,2} = \[-{\Delta\overline{q}\over 4k}\pm\sqrt{\({\Delta\overline{q}\over 4k}\)^2+{\Delta\overline{b}\over 2k}\;}\]_{1,2}\, .
\label{eq:SIDR2int}
\ee 
Here $\hat{c}^{\pm}_{1,2}$ is the intrinsic phase speed of the 4 waves (two at each interface) {\it in the absence of interaction}. To avoid confusion we emphasize that 
$\[\dot{\theta}^{\pm}/k -\ol{U}\]_{1,2} \neq \hat{c}^{\pm}_{1,2}$, since ${\theta}^{\pm}_{1,2}$ is affected by the interaction at a distance with the waves at the opposed interface. All fields of each wave propagate in concert with the same instantaneous frequency ${\dot \theta}^{\pm}_{1,2}$.
Similarly the vorticity and the displacement wave amplitudes change in time due to this interaction, however since in this partition the {\it waves' structures are preserved}, the ratio 
 between vorticity and displacement, $\(2k\hat{c}^\pm\)_{1,2}$, always remains constant (either with or without interaction with the opposed interfacial waves). We therefore refer to these waves as the ``building blocks'' of the linearized interfacial dynamics. While the structure of each wave is untilted (for instance, their far field cross-stream velocity $w$ remains untilted, as can be verified from (\ref{eqwqG}, \ref{eqG})) their superposition may yield a complex tilted structure.

We  immediately note that PM preserves the same simple structure of (\ref{eq:sinpe}):
\be 
{\cal P}^{\pm}_{1,2} = \[\pm {1\over 2} (Z^{\pm})^2\sqrt{\({\Delta\overline{q}\over 2}\)^2+2k\Delta\overline{b}\;}\]_{1,2}\, ; \hspace{0.25cm}  
{\cal P} = \sum_{n=1}^{2} \({\cal P}^{+} +{\cal P}^{-}\)_n.
\label{eq:sinpe2int}
\ee
The energy (\ref{eq:En}) however includes mixed terms between the interfaces since the streamfunction at each interface is contributed both from the waves {\it in-situ} and from the waves of the opposed interface. Using the Green function (\ref{eqG}) we find that  
\be
\nonumber
{\cal E} = 
{k\over 2}\sum_{n=1}^{2} 
\[\(\(\hat{c}^{+}\)^{2} + {\Delta \ol{b} \over 2k}\)
\(Z^+\)^2\ + \(\(\hat{c}^{-}\)^{2} + {\Delta \ol{b} \over 2k}\) \(Z^-\)^2\]_n +
\ee 
\be 
\nonumber
k e^{-k|\Delta z|}\left[{\hat c}_1^+ {\hat c}_2^+ Z_1^{+} Z_2^{+}\cos{(\theta_2^+ -\theta_1^+)} +                       {\hat c}_1^- {\hat c}_2^- Z_1^{-} Z_2^{-}\cos{(\theta_2^- -\theta_1^-)}\, + \right.
\ee
\be
\hspace{0.5cm} \left. {\hat c}_1^+ {\hat c}_2^- Z_1^{+} Z_2^{-}\cos{(\theta_2^- -\theta_1^+)} +                       {\hat c}_1^- {\hat c}_2^+ Z_1^{-} Z_2^{+}\cos{(\theta_2^+ -\theta_1^-)}\right]\, . 
\label{eq:energy2int}
\ee
We write the terms in (\ref{eq:energy2int}) symbolically as
\be 
{\cal E} = \sum_{n=1}^{2}\( {\cal E}^{+} + 
{\cal E}^{-}\)_n + \
\sum_{i=1}^{2}\sum_{j=1}^{2}\({\cal E}^{++}+{\cal E}^{--} +{\cal E}^{+-}+{\cal E}^{-+}\)_{i,j (i\neq j)},
\label{eq:symenergy2int}
\ee
where the first sum includes the waves' self contributions to the energy and the second double sum includes the mixed contributions. We define the interfacial WA (only for the self contribution energy terms) as
\be 
{\cal A}^{\pm}_{1,2} \equiv \({{\cal E} \over\hat{\omega}}\)^{\pm}_{1,2} = 
\[\(\hat{c}^{\pm} + {\Delta \ol{b} \over 2k{\hat c}^{\pm}}\){\(Z^{\pm}\)^2\over 2 }\]_{1,2} = 
 {{\cal P}^{\pm}_{1,2} \over k}\, ,
\label{eq:wa2int}
\ee
so that (\ref{eq:sinwa}) holds. Then we can write PE as
\be 
{\cal H} =  k\sum_{n=1}^{2} \[\({\ol U}+ {\hat c}^+ \){\cal A}^{+} +\({\ol U}+ {\hat c}^- \){\cal A}^{-}\]_n + \sum_{i=1}^{2}\sum_{j=1}^{2}\({\cal E}^{++}+{\cal E}^{--} +{\cal E}^{+-}+{\cal E}^{-+}\)_{i,j (i\neq j)}\, . 
\label{eq:pe2int}
\ee

\subsection{Wave interaction equations}

The following derivation is based on \cite{harnik2008}, hereafter referred to as H08. For clarity we re-derive the essence of it with the notation of the current paper.

We wish to describe the wave interaction dynamics solely in terms of the waves' displacements at the interfaces. 
Toward this end we first take the vorticity equation (\ref{eqq}) and use (\ref{eq:eigen_m2int}, \ref{eq:qzeta2int}) to write
\be 
\[ \hat{c}^+ {D \zeta^+\over Dt} + \hat{c}^- {D \zeta^-\over Dt} =  -{1\over 2k}\(\Delta\overline{q}w +\ii k \Delta\overline{b}\zeta\)\]_{1,2}.
\label{eq:vort2int}
\ee
After this we implement the kinematic condition at the interfaces: 
\be 
\[ {D \over Dt}\(\zeta^+ +  \zeta^- \) = w\]_{1,2}\, , 
\label{eq:disp2int}
\ee
and express the vertical velocity using (\ref{eqwqG}, \ref{eqG}, \ref{eq:qzeta2int}):
\be 
w_{1,2} = -{\ii \over 2}\({\tilde q}_{1,2} + {\tilde q}_{2,1}e^{-k|\Delta z|}  \) = 
-\ii k\[\(\hat{c}^+\zeta^+ + \hat{c}^- \zeta^-  \)_{1,2} + \(\hat{c}^+\zeta^+ + \hat{c}^- \zeta^-   \)_{2,1}e^{-k|\Delta z|}   \]\, .
\label{eq:w2int}
\ee
Then we substitute (\ref{eq:w2int}) in (\ref{eq:vort2int}, \ref{eq:disp2int}), and  after some algebra obtain the equations for the time variation of displacement of each interfacial wave: 
\be  
\dot{\zeta}^{\pm}_{1,2} = -\ii k\[\(\ol{U}+{\hat c}\)_{1,2}{\zeta}^{\pm}_{1,2}\,
\pm e^{-k|\Delta z|}\({\hat{c}^{\pm}\over \hat{c}^+ - \hat{c}^- } \)_{1,2}
\(\hat{c}^+\zeta^+ + \hat{c}^- \zeta^-  \)_{2,1}\]\, , 
\label{zetadot2int}
\ee
where the first term in the RHS is due to the self interfacial wave dynamics and the second results from the interaction at a distance with the two waves at the opposed interface.
Taking the imaginary part of (\ref{zetadot2int}) we get
\be 
\nonumber
\dot{\theta}^{\pm}_{1,2} = k\(\ol{U}+{\hat c}\)^{\pm}_{1,2} \pm
\ee
\be 
{k e^{-k|\Delta z|} \over Z^{\pm}_{1,2}}\({\hat{c}^{\pm}\over \hat{c}^+ - \hat{c}^- } \)_{1,2}
\[\(\hat{c}^+Z^+\)_{2,1}\cos{(\theta_{2,1}^+ -\theta^{\pm}_{1,2})} + \(\hat{c}^- Z^-  \)_{2,1}\cos{(\theta_{2,1}^- -\theta^{\pm}_{1,2})} \].
\label{theta_dot2int}
\ee
The first term in the RHS contains the advection of the mean flow at the interface and the self propagation mechanism in the absence of interaction. The last two terms represent the interaction with the two waves at the opposed interface. The dependence of the interaction on the cosine of their phase difference indicates  the mechanism of interaction. When the waves are in (out of) phase (the cosine is $1$ ($-1$)), the self and the induced cross-stream velocity are in (out of) phase, hence the waves help (hinder) each other to propagate (for more details the reader is kindly referred to the review paper \cite{carp2012}).  

Next we multiply each of the four equations, represented by (\ref{theta_dot2int}), with their counterparts at (\ref{eq:wa2int}) and compare their product with (\ref{eq:pe2int}). After some algebra we obtain the remarkable result:
\be 
\sum_{n=1}^2\[\(\dot{\theta}{\cal A}\)^+ + \(\dot{\theta}{\cal A}\)^-\]_n  = {\cal H}\, ,
\label{Hamilton2int}
\ee
from which
\be 
\der{\cal{H}}{{\cal A}^{\pm}}_{1,2} = \dot{\theta}_{1,2}^{\pm}\, .
\label{1Hamiltoneq2int}
\ee
Finally, taking the real part of (\ref{zetadot2int}),  multiplying it by 
$\[\(\hat{c}^{\pm} + {\Delta \ol{b} \over 2k{\hat c}^{\pm}}\)Z^{\pm}\]_{1,2}$, and noting that 
$\[\(\hat{c}^{\pm} + {\Delta \ol{b} \over 2k{\hat c}^{\pm}}\)= \pm\(\hat{c}^+ - \hat{c}^- \)\]_{1,2}$,
we obtain
\be 
\dot{\cal A}^{\pm}_{1,2} =
-ke^{-k|\Delta z|}\hat{c}^{\pm}_{1,2}
\[\(\hat{c}^+Z^+\)_{2,1}\sin{(\theta_{2,1}^+ -\theta^{\pm}_{1,2})} + \(\hat{c}^- Z^-  \)_{2,1}\sin{(\theta_{2,1}^- -\theta^{\pm}_{1,2})} \]Z^{\pm}_{1,2}.
\label{A_dot2int}
\ee
The above equation indicates that the growth of WA of each interfacial wave is solely due to the interaction with the opposed interfacial waves. The dependence of the interaction on the sine of their phase difference results from the mechanism of growth - the induced cross-stream velocity amplify the wave displacement of the opposed wave and this amplification is maximized when the waves are in quadrature (as in figure \ref{fig:2abcd}(d)). For more details the reader is referred again to \cite{carp2012}.

If we now differentiate (\ref{eq:pe2int}) with respect to ${\theta}_{1,2}^{\pm}$ and compare with (\ref{A_dot2int}), after some algebra we indeed find that
\be 
\der{\cal H}{{\theta}^{\pm}}_{1,2} = -\dot{\cal A}_{1,2}^{\pm}\, .
\label{1Hamiltoneq2int}
\ee
This completes the explicit derivation of the generalized A-A formulation for two interfaces. The procedure can be carried on systematically (not shown here) for multiple interfaces to obtain (\ref{eq:multAA}) for any $N$ integer number of interfaces. 

\subsection{The subset of counter-propagating waves}

As indicated from the heuristic arguments in the introduction, we expect that the instability mechanism will be obtained mainly through action at a distance between the counter-propagating interfacial vorticity waves. Indeed, \cite{rabinovich2011} analyzed the Taylor-Caulfield instability \cite[]{caul1994} and showed that for a large range of Richardson numbers the growth rates, corresponding to the most unstable modes, are practically unaffected when the pro-propagating waves are neglected. Similar results have been obtained by \cite[]{heifetzstratified2015} for the more complex non-Boussinesq dynamics of stratified shear flows \cite[]{heifetzstratified2015}, for swirling flow instability in rotating cylinders \cite[]{RonYellin2017}, for gravity-capillary waves \citep{biancofiore2017}), and even for Alfv\'en waves in magneto-hydrodynamic shear flows \cite[]{heifetzalfven2015}. 
Therefore, next we consider the subset dynamics of the counter propagating interfacial vorticity waves. 

First we need to identify the counter-propagating waves. We note that the conservation of pseudo-momentum and pseudo-energy are unaffected by Galilean transformation.
Hence in the frame of reference of the the mean zonal mean velocity $\ol{U}_m \equiv \(\ol{U}_1 + \ol{U}_2\)/2$ the new Hamiltonian 
${\cal H}_m \equiv \< E + \(\ol{U} - \ol{U}_m \)P\>$ is also a constant of motion (since both $\<P\>$  and $\ol{U}_m$ are constant). Define the zonal mean flow at the reference frame of $\ol{U}_m$ as
$\ol{U}^*_{1,2} \equiv \(\ol{U}_{1,2} - \ol{U}_m\)= \(\ol{U}_{2,1} - \ol{U}_{1,2}\)/2$, we define the counter-propagating waves as the ones whose intrinsic phase speed has the opposite sign of $\ol{U}^*$ at their interface. For instance, let us assume that the shear at the two interfaces is as in figure 2 (level 1 is below level 2) so that the counter-propagating waves will be the ones of figure 2(d), i.e.,
$(\zeta_1^+, \zeta_2^-)$.

Equation (\ref{A_dot2int}) indicates immediately that even if we begin with a pair of counter-propagating waves, the pro-propagating ones $(\zeta_1^-, \zeta_2^+)$ will be generated immediately due to the interaction. Hence the counter-propagating wave subset is just an approximation to the dynamics which is valid only when the waves' amplitudes satisfy $Z_{pro} << Z_{counter}$. Under this approximation the energy and PE become
\be
\nonumber
{\cal E} = 
{k\over 2} 
\[\(\(\hat{c}^{+}\)^{2} + {\Delta \ol{b} \over 2k}\)
\(Z^+\)^2\]_1 + {k\over 2} 
\[\(\(\hat{c}^{-}\)^{2} + {\Delta \ol{b} \over 2k}\) \(Z^-\)^2\]_2 +
\ee 
\be
\hspace{0.5cm} 
k e^{-k|\Delta z|} {\hat c}_1^+ {\hat c}_2^- Z_1^{+} Z_2^{-}\cos{(\theta_2^- -\theta_1^+)} ,              
\label{eq:energy2intcounter}
\ee
%\newpage
\be 
\nonumber
{\cal H} =  k \[\({\ol U}+ {\hat c}^+ \){\cal A}^{+}\]_1 +k \[\({\ol U}+ {\hat c}^- \){\cal A}^{-}\]_2 +  
\label{eq:pe2intcounter}
\ee
\be
\hspace{0.5cm} 
k e^{-k|\Delta z|} {\hat c}_1^+ {\hat c}_2^- Z_1^{+} Z_2^{-}\cos{(\theta_2^- -\theta_1^+)} .              
\label{eq:pe2intcounter}
\ee
The counter-propagating wave interaction equations become
\be 
\dot{\theta}^{+}_{1} = k\(\ol{U}+{\hat c}\)^{+}_{1} +
{k\over 2}  e^{-k|\Delta z|} 
\hat{c}^{+}_1 \hat{c}^{-}_2 {Z_1^+ Z^{-}_{2}\over {\cal A}^{+}_{1}}
\cos{(\theta_{2}^- -\theta^{+}_{1})}\, ,
\label{theta_dot2intcounter1}
\ee
\be 
\dot{\theta}^{-}_{2} = k\(\ol{U}+{\hat c}\)^{-}_{2} +
{k\over 2} e^{-k|\Delta z|} 
\hat{c}^{+}_1 \hat{c}^{-}_2 {Z_1^+ Z^{-}_{2}\over {\cal A}^{-}_{2}}  
\cos{(\theta_{2}^- -\theta^{+}_{1})}\, ,
\label{theta_dot2intcounter2}
\ee
\be 
\dot{\cal A}^{+}_{1} =
-ke^{-k|\Delta z|}\hat{c}^{+}_{1}
\hat{c}^-_2 Z^+_1 Z^-_2\sin{(\theta_{2}^- -\theta^{+}_{1})} =   -\dot{\cal A}^{-}_{2} \, ,
\label{A_dot2intcounter}
\ee
from which it is clear that
\be 
\(\dot{\theta}{\cal A}\)^+_1 + \(\dot{\theta}{\cal A}\)^-_2  = {\cal H}\, ,
\label{Hamilton2intcounter}
\ee
\be 
\der{\cal{H}}{{\cal A}^{+}}_{1} = \dot{\theta}_{1}^{+}\, , 
\hspace{0.5cm}
\der{\cal{H}}{{\cal A}^{-}}_{2} = \dot{\theta}_{2}^{-}\,
\label{1Hamiltoneq2intcounter}
\ee
\be 
\der{\cal H}{{\theta}^{+}}_{1} = -\dot{\cal A}_{1}^{+}\, ,
\hspace{0.5cm}
\der{\cal H}{{\theta}^{-}}_{2} = -\dot{\cal A}_{2}^{-}\, .
\label{1Hamiltoneq2intcounter}
\ee
The same procedure can be carried on naturally for any number ($\geq 2$) of interfaces, where the counter-propagating wave at interface $n$ is the one whose intrinsic phase speed ${\hat c}_n$ is of opposite sign to $\ol{U}^*_n = \(\ol{U} - \ol{U}_m\)_n$, 
$\ol{U}_m$ being the mean zonal mean velocity at all of the interfaces. 
This counter-propagation interaction approximation reduces the  complexity of the dynamics by a factor of two. As shown by \cite{rabinovich2011} and \cite{carpenter2017}, a dense enough grid of interfaces can accurately resolve complex shear flow dynamics, including the rapid changes across critical layers.

\section{Discussion}
\label{sec:5}

In classical mechanics, the action-angle formulation is usually implemented to solve integrable systems, where the action of each degree of freedom (d.o.f) is defined by integrating its generalized momentum around a closed path in the canonical phase space coordinates. For rotating or oscillating conservative motion, the action of each d.o.f is individually conserved, and the total energy is the sum of the product between the action and the intrinsic frequency of all d.o.fs in the system. 
In shear flow the conservation of wave-action is usually exploited to understand the energy change of rays propagating in oblique directions to the mean flow. The propagation component across the shear
alters the Doppler shift felt by the ray which changes its intrinsic frequency and consequently its energy. 

Here we considered an approach, denoted by us as ``generalized action-angle'' dynamics,  which contains mixed components of the latter two. Somewhat similar to a coupled system of $N$ harmonic oscillators we discretize the linearized shear flow dynamics to $M$ interfaces, where on each interface there exists two interfacial waves (so that the number of d.o.f is $N=2M$). Each wave is untilted and propagates in the zonal direction either with the local mean flow or against it. In the absence of interaction (a single interface), it propagates with its intrinsic frequency, with  respect to the mean flow at the interface, and with a constant amplitude and hence a constant action (similar to a single uncoupled oscillator). In  presence of multiple interfaces, the waves interact at a distance by inducing their cross-stream velocity on the remote interfaces, which affects both the waves phases and amplitudes. Therefore changes in the wave frequency is not due to propagation across the shear as in ray tracing dynamics but rather from ``helping or hindering'' the waves at a distance to propagate in the zonal direction. 
The changes in the waves' amplitudes reflects changes in the wave-action. Thus, as opposed to classical systems, the action of each wave is not conserved and this allows instability or transient growth. It is  the sum of the wave action contributions of all waves that is conserved, and is proportional to the domain integrated pseudo-momentum, which is a constant of motion of the linearized system (the linearization allows as well the Fourier decomposition and the consideration of each zonal wavenumber separately). The linearized system as a whole conserves the pseudo-energy, which serves as the Hamiltonian of this generalized action-angle formulation. We emphasize here that in our generalized formulation, the properties of  classical action-angle formulation (that the action-angle variables define an invariant torus and leads to an integrable dynamical system, as mandated by the theorem of Liouville and Arnol'd \citep{arnol2013}) are \emph{not} applicable in general. The reason is simply because individual wave's action is not constant. However (\ref{eq:multAA}) shows that the Hamiltonian $\cal H$ is independent of $\theta$, making the latter a cyclic or angle coordinate.

This work is part of an attempt to develop a compact canonical Hamiltonian theory for shear instability which is both mechanistically intuitive and rigorous. It relies on the mechanistic explanation of Rossby wave instability of \cite{hosk1985} for shear flows that materially conserve vorticity (or potential vorticity) and on the canonical Hamiltonian formulation of it by  \cite{heifetz2009canonical}. The generalization of the mechanistic picture to stratified (non-conserved vorticity) shear flows by \cite{harnik2008}, based on the studies of \cite{holm1962,bain1994,caul1994}, suggested that the essence of the instability remains as a resonant interaction between counter-propagating vorticity waves.
Therefore in this paper we have stressed the link between  action-angle formulation, the conservation laws of pseudo-momentum and pseudo-energy, and the necessary conditions for maintaining wave resonance instability. 

Indeed we can isolate the counter-propagating vorticity waves from the pro-propagating ones by moving to  the frame of reference of the mean zonal mean velocity $\ol{U}_m$. The conservation of pseudo-momentum and pseudo-energy are unaffected by Galilean transformation, hence the new Hamiltonian 
${\cal H}_m \equiv \< E + \(\ol{U} - \ol{U}_m \)P\>$ is also a constant of motion. 
Hence we can define the counter-propagating waves in the frame of reference of $\ol{U}_m$ as $P^-$ for 
$\(\ol{U} - \ol{U}_m \) > 0$ and $P^+$ for $\(\ol{U} - \ol{U}_m \) < 0$. 
If we choose to neglect the pro-propagating waves in these layers, we lose accuracy as we reduce the system's d.o.f from $2M$ to $M$. Nevertheless it can be shown that the reduced subset
system still preserves the action-angle formulation. \cite{rabinovich2011} analyzed the Taylor-Caulfield instability and showed that for a large range of Richardson numbers, the growth rates corresponding to the most unstable modes are
practically unaffected if the pro-propagating waves are neglected. Furthermore, they managed to resolve the critical layer dynamics of continuous profiles, with sufficient accuracy, when implementing the interfacial wave dynamics with a high resolution gird. 

The vorticity wave interaction in stratified shear flows has been generalized further to include non-Boussinesq effects \cite[]{heifetzstratified2015}, as well as surface tension between immiscible fluids \cite[]{biancofiore2015}. It has been implemented for swirling flow instability in rotating cylinders  \cite[]{RonYellin2017}, and even for Alfv\'en waves in magneto-hydrodynamic shear flows \cite[]{heifetzalfven2015}. Therefore, it is our plan to generalize this generalized action-angle formulation further to such more complex setups of shear flows. 

\appendix
\section{Relation of pseudo-energy conservation with the Howard-Miles criterion for instability}\label{App:A}

\setcounter{equation}{0}

\renewcommand{\theequation}{A.\arabic{equation}}

The seminal Howard-Miles criterion states that a necessary condition for modal instability (of the form of $e^{ik(x-ct)}$ with  $c_i > 0$) is that the Richardson number, $Ri(z) \equiv \ol{b}_z/ (\ol{U}_z)^2$, should be smaller than a quarter somewhere within the domain \citep{draz1982}. For modal instability the PE must be zero,
$\der{}{t}{\cal H} = 2kc_i{\cal H}=0$, thus 
\be  
\<E\> = -\<\ol{U}{\it P}\> .   
\label{Hzero}
\ee
Writing the velocity and the cross-stream displacement in terms of the streamfunction $\psi$:
$u= -\der{\psi}{z}$, $w = \der{\psi}{x} = \ii k\psi$, 
${D\zeta \over Dt} = w\, \Rightarrow \zeta = {\psi / (\ol{U} - c)}$, and substitute in (\ref{eq:En})
we obtain
\be 
\<E\> = {1\over 2}\<k^2|\psi|^2 +  |\der{\psi}{z}|^2 + {\ol{b}_z \over |\ol{U} - c|^2}|\psi|^2\>.
\ee
Now, in the original paper by \cite{miles1961}, and its generalization by \cite{howard1961}, the Taylor-Goldstein equation has been manipulated and integrated by parts (see Chapter 7 of \cite{kundu2004} for explicit derivation) to finally obtain
\be 
c_i\<E\> = c_i\[{1\over 2}\<{{1 \over 4} (\ol{U}_z)^2 \over |\ol{U} - c|^2}|\psi|^2\>\]
\label{Howard}
\ee
from which Howard concluded that $c_i \neq 0 $ is possible only if $Ri < 1/4$ somewhere within the domain.
Thus, contrast (\ref{Howard})  with (\ref{Hzero}) we see that for $c_i \neq 0 $:
\be 
\<\ol{U}{\it P}\> = -{1\over 2}\<{{1 \over 4} (\ol{U}_z)^2\over |\ol{U} - c|^2}|\psi|^2\>.
\ee
We note that it seems that neither Miles, nor Howard, have related those integrals to their physical interpretation as wave energy and pseudo-energy.

\bibliographystyle{jfm}
\bibliography{references}

\end{document}